\documentclass[useAMS,usenatbib]{mn2e}
\usepackage{graphicx}

\title[Suppression of hose instability]
{Suppression of resistive hose instability in a
relativistic electron--positron flow}
\author[M.~Honda]{Mitsuru Honda\\
Plasma Astrophysics Laboratory, Institute for Global Science,
Mie 519-5203, Japan}
\begin{document}

\date{Accepted 2007 January 8. Received 2006 December 2;
in original form 2006 March 6}

\pagerange{\pageref{firstpage}--\pageref{lastpage}} \pubyear{2007}

\maketitle

\label{firstpage}

\begin{abstract}
This paper presents the effects of electron--positron pair production
on the linear growth of the resistive hose instability of a
filamentary beam that could lead to snake-like distortion.
For both the rectangular radial density profile and
the diffuse profile reflecting the Bennett-type
equilibrium for a self-collimating flow, the modified
eigenvalue equations are derived from a Vlasov--Maxwell equation.
While for the simple rectangular profile, current perturbation is
localized at the sharp radial edge, for the realistic Bennett
profile with an obscure edge, it is non-locally distributed over the
entire beam, removing catastrophic wave--particle resonance.
The pair production effects likely decrease the betatron frequency, and
expand the beam radius to increase the resistive decay time of the
perturbed current; these also lead to a reduction of the growth rate.
It is shown that, for the Bennett profile case, the characteristic
growth distance for a preferential mode can exceed
the observational length-scale of astrophysical jets.
This might provide the key to the problem of the stabilized
transport of the astrophysical jets including extragalactic
jets up to Mpc ($\sim 3\times 10^{24}~{\rm cm}$) scales.
\end{abstract}


\section{INTRODUCTION}

Supra-parsec-scale transport of astrophysical jets has been
a puzzle for a long time \citep[e.g.][]{bridle84}, and
substantial efforts have been devoted, both theoretically
and observationally, to solve their extremely stable
feature \citep[for a review, see][]{hardee04}.
According to the first principle analysis \citep{appl92},
the outflows driven by a central engine are expected to
carry a huge current that could reach $10^{17}~{\rm A}$
and more \citep[for the jet of 3C\,273, see, for example,][]{conway93}.
However, it is known that reflecting the interplay between
the beam and the self-generated azimuthal magnetic field,
the net current is inhibited \citep{alfven39}.
The upper limit associated with a unit current
is given by \citep{honda00}
\begin{eqnarray}
I_{0}\simeq 28\beta\Gamma~{\rm kA},
\label{eq:1}
\end{eqnarray}
\noindent
where $\Gamma=(1-\beta^{2})^{-1/2}$ is the Lorentz factor
of an electron (or positron) flow.
Apparently, here there is the serious difficulty that
the current of jets greatly exceeds $I_{0}$ for
conceivable $\Gamma$-values.
One possible idea to overcome this is to make the
cluster of current filaments, each having a current
below $I_{0}$, carry such a huge current \citep{hh02}.
This situation might be accomplished via the current
filamentation instability (CFI) including the Weibel mode
\citep[cf.][and references therein]{honda04}.
Indeed, the perfect conductivity characteristic of plasma
calls the total return current equal to the outgoing current,
likely arranging a pattern of the counterstreaming currents,
which could serve as a free-energy source to amplify
the perturbed magnetic fields transverse to the currents.
In an environment with body-wave perturbations unstable
for such an electromagnetic mode, a two-stream
instability for longitudinal electrostatic perturbations
\citep[e.g.][]{buneman58} could simultaneously develop,
although this is considered not to play an essential role in
establishing the clustered current system \citep[e.g.][]{nishikawa03}.
Also, highly self-organized filaments are
certainly confirmed in many radio sources,
for example the Galactic Centre
\citep{yusefzadeh84,yusefzadeh87,yusefzadeh04},
Cyg\,A \citep{perley84}, M87 \citep{owen89},
3C\,353 \citep{swain98}, 3C\,273 \citep{lobanov01}
and 3C\,438 \citep{treichel01}.
Moreover, it has recently been suggested that the
filamentary jet model \citep{hh04,hh05} could reproduce
the synchrotron X-ray spectra observed
in extragalactic jets \citep{fleishman06,hh07}.

Long-scale jets have narrow opening angles less than
$10\degr$, although in close proximity to the central
engine the angles tend to significantly spread
\citep[e.g. $\sim 60\degr$ for the M87 jet;][]{junor99}.
In addition, there is a consensus that the internal
pressure of the jets is higher than the external pressure:
for example, 4C\,32.69 \citep{potash80}, Cyg\,A
\citep{perley84} and M87 \citep{owen89}.
It is implied that the jets are self-collimating.
In a promising case where the outgoing currents are collimated
by the self-generated magnetic fields, the magnetic pressure
exerts to preferentially evacuate the plasma return currents
radially outward \citep{honda00a,honda00b}.
The resulting radial expansion might be pronounced
for the plasma around the envelope of the filament
cluster, because of the lower external pressure.
As shown by the observational facts, a portion of the
evacuated flows can constitute a halo surrounding jet
\citep[e.g. 1803+784;][]{gabuzda03,britzen05},
filamentary back streams \citep[3C\,84;][]{asada06},
a cocoon (Mrk\,3, \citealt{capetti99}; 3C\,273,
\citealt{bahcall95}; Cyg\,A, \citealt{carilli96}), and so on.
Importantly, as a result of the continuous evacuation,
the vacuum regions could appear to mediate the beam
and return currents \citep{honda00a,honda00b}: the
velocity shear-free configuration is likely established.
Then, the surface mode instability driven by the excess
kinetic energy of the shear flows \citep[i.e. the
Kelvin--Helmholtz instability in a fluid context; e.g.][]
{frank96,hanasz96,malagoli96} is not crucial.
However, if the fractional return current remains in the beam core,
the instability driven by the counterstreaming currents, such as
the hose instability for kink-type perturbations, can grow.
The eigenfunction is peaked near the surface, involving,
for example, for a simple beam with rectangular radial density
profile, the collective resonance with the beam electrons
undergoing betatron oscillation (cf. Section 5.1).
In this sense, the hose instability might also be
classified into the surface mode instability.
The bulk mode instabilities leading to sausage- or hollowing-like
deformation of beam \citep{uhm81,joyce83} are ignored
here, to isolate the axially asymmetric modulation.

The beam kinetic energy tends to be, in part, converted to
transverse thermal energy via collisionless processes \citep{honda00b},
whereupon the pressure equilibrium is hydrodynamically determined
by the transverse dynamics \citep{honda00,honda00a}.
However, the conventional magnetohydrodynamic (MHD) description
\citep[e.g.][]{chandrasekhar81} is insufficient for the stability
analysis, as it averages out the betatron oscillation of beam
particles, merely remaining an averaged flow velocity.
If any, the cold return plasma that comprises
gyrating particles can be modelled as an MHD fluid
(including the negligible effects of particle inertia).
However, note that, in realistic situations, the energy flux
of beams is expected to be larger than that of the
return flows, such that the beam dynamics dominantly
influences the evolution of the entire system.
As a consequence, for a collisionless, isolating beam
(surrounded by the vacuum) the Vlasov equation is found to be
adequate for resolving the wave--beam particle interaction.
In a fluid-like way, the beam column appears to be
stable for the velocity shear-free configuration.
It is also noted that macroscopic dynamics of the cluster of
filaments could deviate from that of a uniformly filled
cylinder described by the fluid equation, as the filaments
are electromagnetically decoupled, retaining the coherence of
the betatron orbital motion (reflected in the beam structure).
According to the kinetic simulations of asymmetric counterstreaming
currents (reflecting the energy flux budget mentioned above),
indeed, the mutual coupling of filaments seems likely to be
poor in the fully developed non-linear phase of the CFI
(except for a peculiar phase of rapid coalescence of filaments),
owing to the return current sheath \citep{honda00a}.
In another regime in which filamentary turbulence
is well correlated all over the system (larger than the
amplitude of the betatron oscillation and the gyroradius),
the kinetic analysis could provide a microscopic basis
for the macroscopic fluid approach particularly involving
an anomalous resistivity for the filamentary medium \citep{hh05}.
However, one should notice the crucial point that the
collective wave--particle resonance possibly gives rise
to a morphologic catastrophe in the non-linear phase;
this is sheer unpredictable in the fluid context.

In this aspect, the recurrence of the pioneering kinetic
analysis by \citet{uhm80} may be useful for studying the
fundamental transport property, although their argument
was limited to be of a stability check of the electron
beam (without positrons) produced in the laboratory.
The key consequence is that for the self-collimating electron
beam with a diffuse radial density profile reflecting the so-called
Bennett pinch equilibrium \citep{bennett34}, the resonant
divergence of a growth rate is removed, in contrast to the
aforementioned simple case with a rectangular profile.
Related to this, the kinetic simulations
\citep{honda00a} reproduced the self-organization
of a Bennett-like profile \citep{honda00}.
Thus, the stable (or quasi-stable) transport of
self-collimating flows is anticipated to be
realized in various situations exhibiting similarity.
In particular, for an application of their analysis to astrophysical
jets, it is very important to take account of the positron
abundance effects \citep[e.g.][]{roland95} -- BL\,Lacs \citep{xie95},
M87 \citep{reynolds96} and 3C\,279 \citep{wardle98} --
whose details have not substantially been investigated so far.

In this paper, the Vlasov--Maxwell analysis of resistive hose
instability, which was briefly introduced in \citet{hh02}, is
expanded to systematically survey the possible unstable modes
involved in the relativistic electron--positron flows with a
Bennett-type profile and, for comparison, the rectangular profile.
In the astrophysical context, the focus here is on highlighting
some noticeable features of pair production effects, rather than
a parameter survey such as was performed by \citet{uhm80}.
It is shown that the charge screening effects by
positrons directly lower the betatron frequency to
suppress the linear growth of the convective mode.
The lowered betatron frequency appears to be reflected in the
expansion of the filament radius, which prolongs the resistive
decay time of the perturbed current, to suppress the growth of
the absolute mode; whereupon for the Bennett profile case,
the spatial growth distance can be safely longer than (or
comparable to) the observational length-scale of astrophysical jets.
For this case, for convenience, the scaling of the growth distance
in the relevant plasma-parameter region is explicitly shown.

In order to present these details in a straightforward way,
this paper has been organized as follows.
In Section~2, an ad hoc model of negatively charged
electron--positron gas is introduced.
For the relativistic flows, in Section~3 the equilibria of the
momentum distribution functions and self-generated magnetic
fields are constructed for the cases of the rectangular density
profile (Section~3.1) and Bennett profile (Section~3.2).
Then, in Section~4, the Vlasov--Maxwell equations
are linearized around the equilibria.
Using the procedure outlined in Appendix, the eigenvalue equations
are derived from the linearized equation, to extract the relevant
eigenmodes in Section~5, for the cases of the rectangular
(Section~5.1) and Bennett (Section~5.2) profiles.
Section~6 is devoted to a discussion concerning mode preference
in actual circumstances, and a summary.

\section{A SIMPLE MODEL OF ELECTRON--POSITRON GAS}

The extreme astrophysical environments, in which
electron--positron pairs are created, can be typically
found in galactic nuclei (e.g. \citealt{zdziarski85,heyl01};
for a laboratory experiment, see \citealt{wilks05}).
As a rule, the charge conservation law requires that the total electron
charge is compensated by the pair-created positrons and discharged ions,
such that $n_{\rm e}=n_{\bar{\rm e}}+\left<Z^{*}\right>n_{\rm i}$,
where $n_{\rm e}$, $n_{\bar{\rm e}}$ and $n_{\rm i}$ are the
total electron, positron and ion densities, respectively, and
$\left<Z^{*}\right>$ is the average of the charge state of a composite.
Hereafter, the superscript '$^-$' indicates the quantity
for 'anti'-electron (i.e. positron).
For simplicity, the ration of
$n_{\bar{\rm e}}/n_{\rm e}\equiv f_{\rm pair}$
is supposed to be constant, where $0\leq f_{\rm pair}<1$.
Then the definition
\begin{eqnarray}
\Delta f_{\rm pair}\equiv 1-f_{\rm pair},
\label{eq:2}
\end{eqnarray}
\noindent
is introduced, 
and the reciprocal, $(\Delta f_{\rm pair})^{-1}$, is referred to as
the pair production rate; these take the values in the ranges of
$0<\Delta f_{\rm pair}\leq 1$ and $(\Delta f_{\rm pair})^{-1}\geq 1$,
respectively.
Because $n_{\rm e}>n_{\bar{\rm e}}$, the electron--positron
gas may be regarded as negatively charged.

As discussed in Section~1, in the short-circuit reflecting
the characteristic of perfect conductivity in plasma,
the return currents are allowed to flow outside the beams,
and even the jet \citep[for modeling, cf.][]{benford78,alfven81}.
For the present purpose, the fractional beam density, defined as
the ratio of the beam electron (positron) density to $n_{\rm e}$
($n_{\bar{\rm e}}$), is assumed to be constant [i.e.
$f_{\rm b}=n_{\rm b}/n_{\rm e}(=n_{\bar{\rm b}}/n_{\bar{\rm e}})
={\rm const}$], whereby the fractional return flow density is given by
$1-f_{\rm b}=n_{\rm r}/n_{\rm e}(=n_{\bar{\rm r}}/n_{\bar{\rm e}})
={\rm const}$.
Then we have $n_{\bar{\rm b}}=f_{\rm pair}n_{\rm b}$
and $n_{\bar{\rm r}}=f_{\rm pair}n_{\rm r}$.
Hereafter, the subscripts '${\rm b}$' and '${\rm r}$' indicate
the quantities for the beam and return flow, respectively.
It is also mentioned that, for $(\Delta f_{\rm pair})^{-1}\gg
1840\left<Z^{*}\right>/\left<A^{*}\right>\sim 10^{3}$
and $\ll 10^{3}$, where $\left<A^{*}\right>$ is the average
of mass number, the electron--positron fluid and ion rest frames,
respectively, might be properly referred to as the 'jet frame'.
In order to exclude the ambiguity of the reference, throughout
this paper, all physical quantities, except for the quantity
$\Omega$ introduced in equations~(\ref{eq:29}) and (\ref{eq:30}),
are specified in the ion rest frame \citep{hh02}.

We consider the situation that the electron--positron gas
flowing in the $z$-direction has the speed of $v_{\rm b}$
and the gas flowing in the opposite direction has the speed
of $|v_{\rm r}|$ ($\ll v_{\rm b}$), and these have the
thermal energy of $T_{\rm b}$ and $T_{\rm r}$, respectively.
Then, it is convenient to define the effective parameter of
\begin{equation}
\beta\equiv f_{\rm b}\beta_{\rm b}-(1-f_{\rm b})|\beta_{\rm r}|.
\label{eq:3}
\end{equation}
\noindent
Here, $\beta_{\rm b}=v_{\rm b}/c={\rm const}$ and
$|\beta_{\rm r}|=|v_{\rm r}|/c={\rm const}$,
where $c$ is the speed of light.
Note that equation~(\ref{eq:3}) takes a value
in the range of $0<\beta\leq 1$.
Similarly, the effective temperature is defined as
\begin{equation}
T\equiv f_{\rm b}T_{\rm b}+(1-f_{\rm b})T_{\rm r},
\label{eq:4}
\end{equation}
\noindent
where $T_{\rm b}$, $T_{\rm r}={\rm const}$.
For the special case, $f_{b}=1$, reflecting that the plasma
return current is perfectly evacuated outside the beam core
(Section~1), equations~(\ref{eq:3}) and (\ref{eq:4}) reduce
to $\beta=\beta_{\rm b}$ and $T=T_{\rm b}$, respectively.
The definitions given by equations~(\ref{eq:2})--(\ref{eq:4})
are substantially used throughout.

\section{EQUILIBRIA OF THE RADIALLY CONFINED RELATIVISTIC FLOWS}

In what follows, the zeroth-order distribution functions
of beam electrons are assigned, as they provide, within
the present framework, sufficient information to
characterize the equilibrium properties of the
negatively charged fluid \citep[cf.][]{wilks05}.
First, we consider a simple case of a collimated flow with
thermal spread and sharp radial edge (Section~3.1), and
secondly, another case of a self-collimating flow with
thermal spread and diffuse density profile (Section~3.2).

\subsection{The case of rectangular radial density profile}

Using the similar procedure by \citet{davidson90},
we begin with the equilibrium distribution function
for the beam electrons with the momentum of
${\bf p}=(p_{r},p_{\theta},p_{z})$:
\begin{equation}
f_{\rm b}^{0}({\bf p},r)=
\frac{\hat{n}_{\rm b}}{2\pi\gamma_{\rm b}m_{\rm e}}F_{\rm b}(H)
\delta\left(p_{z}-\gamma_{\rm b}m_{\rm e}\beta_{\rm b}c\right).
\label{eq:5}
\end{equation}
\noindent
Here,
$F_{\rm b}(H)=\delta\left[H
-\left({\hat\gamma_{\rm b}}-1\right)m_{\rm e}c^{2}\right]$,
$\delta$ indicates the Dirac delta function,
$H=(p_{r}^{2}+p_{\theta}^{2})/(2\gamma_{\rm b}m_{\rm e})+
(\gamma_{\rm b}-1)m_{\rm e}c^{2}+e\beta_{\rm b}A_{z}^{\rm s}(r)$
is the Hamiltonian, $e$ is the elementary charge,
$m_{\rm e}$ is the electron rest mass,
$\gamma_{\rm b}=(1-\beta_{\rm b}^{2})^{-1/2}$, and ${\hat\gamma}_{\rm b}$
$(>\gamma_{\rm b})$ includes the thermal component.
In the cylindrically symmetric case, the $z$-component
of the vector potential, $A_{z}^{\rm s}$, conforms
to the $\theta$-component of the magnetic field:
$B_{\theta}^{\rm s}(r)=-{\partial A_{z}^{\rm s}(r)/\partial r}$,
where the superscript '${\rm s}$' indicates the
'self'-generated quantities.
Equation~(\ref{eq:5}) is rewritten as
\begin{eqnarray}
f_{\rm b}^{0}({\bf p},r)=
\frac{\hat{n}_{\rm b}}{2\pi\gamma_{\rm b}m_{\rm e}}
\delta\left[\frac{p_{\perp}^{2}}{2\gamma_{\rm b}m_{\rm e}}
+\psi_{\rm b}(r)
-\left({\hat\gamma_{\rm b}}-\gamma_{\rm b}\right)m_{\rm e}c^{2}\right]
\nonumber \\
\times\delta\left(p_{z}-\gamma_{\rm b}m_{\rm e}\beta_{\rm b}c\right),
\label{eq:6}
\end{eqnarray}
\noindent
where $p_{\perp}^{2}\equiv p_{r}^{2}+p_{\theta}^{2}$, and
$\psi_{\rm b}(r)\equiv e\beta_{\rm b}A_{z}^{\rm s}(r)$ represents
the effective potential that acts on beam electrons.

For the definition of the beam electron density,
$n_{\rm b}(r)\equiv\int{\rm d}^{3}{\bf p}f_{\rm b}^{0}({\bf p},r)
=\pi\int_{0}^{\infty}{\rm d}(p_{\perp}^{2})
\int_{-\infty}^{\infty}{\rm d}p_{z}
f_{\rm b}^{0}(p_{\perp},p_{z},r)$,
we have
\begin{equation}
n_{\rm b}(r)=\hat{n}_{\rm b}\int_{0}^{\infty}{\rm d}U
\delta\left[U+\psi_{\rm b}(r)
-\left({\hat\gamma_{\rm b}}-\gamma_{\rm b}\right)
m_{\rm e}c^{2}\right],
\label{eq:7}
\end{equation}
\noindent
where $U\equiv p_{\perp}^{2}/(2\gamma_{\rm b}m_{\rm e})$.
The integral is equal to unity for
$\psi_{\rm b}(r)<\left({\hat\gamma_{\rm b}}
-\gamma_{\rm b}\right)m_{\rm e}c^{2}$
and equal to zero for
$\psi_{\rm b}(r)>\left({\hat\gamma_{\rm b}}
-\gamma_{\rm b}\right)m_{\rm e}c^{2}$.
Therefore, the density profile has the simple rectangular form
\begin{equation}
n_{\rm b}=\left\{
\begin{array}{ll}
  \hat{n}_{\rm b}={\rm const} & {\rm for}~0\leq r<r_{\rm b} \\
  0 & {\rm for}~r>r_{\rm b}\\
\end{array},
\right.
\label{eq:8}
\end{equation}
\noindent
whereby $n_{\bar{\rm b}}=\hat{n}_{\bar{\rm b}}={\rm const}$ for
$0\leq r<r_{\rm b}$ and $n_{\bar{\rm b}}=0$ for $r>r_{\rm b}$.
The beam filament radius, $r_{\rm b}$ (smaller than the radius
of the jet; Section~1), can be self-consistently determined by
\begin{equation}
\psi(r_{\rm b})=\left({\hat\gamma_{\rm b}}
-\gamma_{\rm b}\right)m_{\rm e}c^{2}.
\label{eq:9}
\end{equation}
\noindent
Note here that the spatial profile of the cylindrical
magnetic field can be expressed as
\begin{equation}
B_{\theta}^{\rm s}(r)=
\left\{
\begin{array}{ll}
  -2\pi e\beta\Delta f_{\rm pair}{\hat n}_{\rm e}r &
  {\rm for}~0\leq r<r_{\rm b} \\
  -2\pi e\beta\Delta f_{\rm pair}{\hat n}_{\rm e}(r_{\rm b}^{2}/r) &
  {\rm for}~r>r_{\rm b}\\
\end{array},
\right.
\label{eq:10}
\end{equation}
\noindent
where ${\hat n}_{\rm e}={\hat n}_{\rm b}/f_{\rm b}$,
and the corresponding effective potential is
$\psi_{\rm b}(r)=\pi e^{2}\beta_{\rm b}
\beta\Delta f_{\rm pair}\hat{n}_{\rm e}r^{2}$
(for $0\leq r<r_{\rm b}$).
Therefore, equation~(\ref{eq:9}) yields
\begin{equation}
r_{\rm b}^{2}=\frac{2c^{2}}{{\hat\omega}_{\beta}^{2}}
\frac{{\hat\gamma}_{\rm b}-\gamma_{\rm b}}{\gamma_{\rm b}},
\label{eq:11}
\end{equation}
\noindent
where ${\hat\omega}_{\beta}(={\rm const})$ defines
the betatron frequency, whose square is given by
\begin{equation}
{\hat\omega}_{\beta}^{2}=\frac{2\pi e^{2}\beta_{\rm b}\beta
\Delta f_{\rm pair}\hat{n}_{\rm e}}{\gamma_{\rm b}m_{\rm e}}.
\label{eq:12}
\end{equation}
\noindent
It is found that the larger pair production rate (i.e.
the smaller value of $\Delta f_{\rm pair}$) leads to
smaller ${\hat\omega}_{\beta}$ and larger $r_{\rm b}$.

\subsection{The Bennett-type equilibrium case with a diffuse
radial density profile}

We consider the radial force balance between the gradient of
static pressure stemming from the thermal spread of the beam,
$\left(\partial/\partial r\right)
\left(1+f_{\rm pair}\right)n_{\rm e}(r)T$, and
the ${\bf J}\times{\bf B}$ contraction force.
This might reflect a more realistic equilibrium
of relativistic electron--positron flow.
When the electrostatic field is negligible \citep{honda00a},
the force balance equation is written as \citep{hh02}
\begin{eqnarray}
T\left(1+f_{\rm pair}\right)\frac{\partial n_{\rm e}(r)}{\partial r}
=-\frac{4\pi e^{2}}{r}\beta^{2}\Delta f_{\rm pair}^{2}n_{\rm e}(r)
\nonumber \\
\times\int_{0}^{r}{\rm d}r^{\prime}r^{\prime}n_{\rm e}(r^{\prime}),
\label{eq:13}
\end{eqnarray}
\noindent
where we use the expression of the cylindrical
magnetic field self-generated as a result of the
net current of an electron--positron flow:
\begin{equation}
B_{\theta}^{\rm s}(r)
=-\frac{4\pi e}{r}\beta\Delta f_{\rm pair}
\int_{0}^{r}{\rm d}r^{\prime}r^{\prime}n_{\rm e}(r^{\prime}).
\label{eq:14}
\end{equation}
\noindent
The spatial profile of the electron number density can be
self-consistently determined, to have the form of
\begin{equation}
\frac{n_{\rm e}(r)}{\hat{n}_{\rm e}}
\left[=\frac{n_{\rm b}(r)}{\hat{n}_{\rm b}}\right]
=\frac{1}{\left(1+r^{2}/r_{\rm b}^{2}\right)^{2}},
\label{eq:15}
\end{equation}
\noindent
where $\hat{n}_{\rm e}\equiv n_{\rm e}(r=0)=\hat{n}_{\rm b}/f_{\rm b}$,
and the characteristic radial size of a beam filament is given by
\begin{equation}
r_{\rm b}^{2}=\frac{2\left(1+f_{\rm pair}\right)T}
{\pi e^{2}\beta^{2}\Delta f_{\rm pair}^{2}\hat{n}_{\rm e}}.
\label{eq:16}
\end{equation}
\noindent
In contrast to equation~(\ref{eq:8}), the density profile
given by equation~(\ref{eq:15}) has no sharp radial edge.
This corresponds to the modified Bennett pinch equilibrium of
a self-collimating relativistic electron--positron flow, in which
spatial profile of the magnetic field can be expressed as
\begin{equation}
B_{\theta}^{\rm s}(r)=-2\pi e\beta\Delta f_{\rm pair}{\hat n}_{\rm e}
\frac{r}{1+r^{2}/r_{\rm b}^{2}}.
\label{eq:17}
\end{equation}
\noindent
Note that $B_{\theta}^{\rm s}(r)$ takes the peak value
at $r=r_{\rm b}$, where the magnetic energy density,
$|B_{\theta}^{\rm s}(r=r_{\rm b})|^{2}/(8\pi)$, is in the
level of $\sim\left(1+f_{\rm pair}\right){\hat n}_{\rm e}T$,
reflecting pressure balance.

Let us consider the equilibrium beam distribution
function of the form of
\begin{eqnarray}
f_{\rm b}^{0}({\bf p},r)=
\frac{\hat{n}_{\rm b}}{2\pi\gamma_{\rm b}m_{\rm e}}
F_{\rm b}\left[\frac{p_{\perp}^{2}}{2\gamma_{\rm b}m_{\rm e}}
+\left(\gamma_{\rm b}-1\right)m_{\rm e}c^{2}
\right. \nonumber \\ \left.
+e\beta_{\rm b}A_{z}^{\rm s}(r)\right]
\delta\left(p_{z}-\gamma_{\rm b}m_{\rm e}\beta_{\rm b}c\right).
\label{eq:18}
\end{eqnarray}
\noindent
Recalling the definitions of $U$ and $\psi_{\rm b}(r)$,
the beam electron density is then given by
\begin{equation}
n_{\rm b}(r)\equiv\hat{n}_{\rm b}\int_{0}^{\infty}{\rm d}U
F_{\rm b}\left[U+\left(\gamma_{\rm b}-1\right)m_{\rm e}c^{2}
+\psi_{\rm b}(r)\right].
\label{eq:19}
\end{equation}
\noindent
Imposing the boundary condition of $A_{z}^{\rm s}(r=0)=0$,
we integrate equation~(\ref{eq:17}), to obtain
$n_{\rm b}(r)/\hat{n}_{\rm b}=\exp\left[-2A_{z}^{\rm s}(r)
/\left(\pi e\beta\Delta f_{\rm pair}\hat{n}_{\rm e}r_{\rm b}^{2}\right)
\right]$,
which is recast to
\begin{equation}
\frac{n_{\rm b}(r)}{\hat{n}_{\rm b}}=\exp\left[
-\frac{\beta\Delta f_{\rm pair}}
{\beta_{\rm b}\left(1+f_{\rm pair}\right)T}\psi_{\rm b}(r)\right].
\label{eq:20}
\end{equation}
\noindent
Here, $\psi_{\rm b}(r)$ [$\equiv e\beta_{\rm b}A_{z}^{\rm s}(r)$]
conforms to the equation of
\begin{equation}
\frac{\partial\psi_{\rm b}(r)}{\partial r}
=\gamma_{\rm b}m_{\rm e}r\omega_{\beta}^{2}(r),
\label{eq:21}
\end{equation}
\noindent
where $\omega_{\beta}$ defines the betatron frequency,
whose square is
\begin{equation}
\omega_{\beta}^{2}(r)=\frac{{\hat\omega}_{\beta}^{2}}
{1+r^{2}/r_{\rm b}^{2}}.
\label{eq:22}
\end{equation}
\noindent
Invoking the density inversion theorem for equation~(\ref{eq:19}),
namely, $F_{\rm b}(H)=-\hat{n}_{\rm b}^{-1}
[\partial n_{\rm b}/\partial\psi_{\rm b}]_{\psi_{\rm b}
=H-(\gamma_{\rm b}-1)m_{\rm e}c^{2}}$,
where $H=U+(\gamma_{\rm b}-1)m_{\rm e}c^{2}+\psi_{\rm b}$
(Section~3.1; \citealt{davidson90}), we explicitly obtain
the expression of $F_{\rm b}(H)$, that is,
\begin{eqnarray}
F_{\rm b}(H)=\frac{\beta\Delta f_{\rm pair}}
{\beta_{\rm b}\left(1+f_{\rm pair}\right)T}
\exp\left\{-\frac{\beta\Delta f_{\rm pair}}
{\beta_{\rm b}\left(1+f_{\rm pair}\right)T}
\right. \nonumber \\ \left.
\times\left[H-(\gamma_{\rm b}-1)m_{\rm e}c^{2}\right]\right\}.
\label{eq:23}
\end{eqnarray}
This constitutes a pseudo distribution function of
thermal equilibrium for equation~(\ref{eq:18}).

If we take account of no direct contribution of return current
component to the thermal pressure gradient and ${\bf J}\times{\bf B}$
force (in the radial force balance), in equations~(\ref{eq:13}) and
(\ref{eq:16}) the replacements of $T\rightarrow f_{\rm b}T_{\rm b}$
and $\beta^{2}\rightarrow f_{\rm b}\beta_{\rm b}\beta$
are valid, whereby in equation~(\ref{eq:23}),
$\beta\Delta f_{\rm pair}/[\beta_{\rm b}\left(1+f_{\rm pair}\right)T]
\rightarrow \Delta f_{\rm pair}/[\left(1+f_{\rm pair}\right)T_{\rm b}]$.
In this regime, for the special case of no pair production
(i.e. $f_{\rm pair}=0$ and $\Delta f_{\rm pair}=1$),
equation~(\ref{eq:23}) reduces to
\begin{equation}
F_{\rm b}(H)=\frac{1}{T_{\rm b}}
\exp\left[-\frac{H-(\gamma_{\rm b}-1)m_{\rm e}c^{2}}{T_{\rm b}}\right],
\label{eq:24}
\end{equation}
which coincides with equation~(9.239) in \citet{davidson90}.

\section{PERTURBED VLASOV--MAXWELL EQUATIONS}

Now we superimpose the fluctuations on the aforementioned
equilibria of $f_{\rm b}^{0}$ and $A_{z}^{\rm s}$, such that
the total distribution function and vector potential can be
linearized as $f_{\rm b}=f_{\rm b}^{0}+\delta f_{\rm b}$
and ${\bf A}=A_{z}^{\rm s}{\hat{\bf z}}+\delta{\bf A}$, respectively.
Assuming a resistive response of return currents, the fluctuating
total current can be phenomenologically expressed as
$\delta{\bf J}=\Delta f_{\rm pair}\delta {\bf J}_{\rm b}
+\sigma\delta{\bf E}$,
where $\delta{\bf J}_{\rm b}$ denotes the fluctuating
beam electron current, $\sigma$ is the electrical conductivity
and $\delta{\bf E}=-(1/c)(\partial\delta{\bf A}/\partial t)$ is
the fluctuating electric field (without electrostatic component;
cf. Section~3).
We also assume a slow spatiotemporal change, such that
$|\omega|\ll 4\pi\sigma$, $|\omega|r_{\rm b}/c\ll 1$
and $|k_{z}|r_{\rm b}\ll 1$ \citep{uhm80,davidson90}.
Then, the linearized Ampere--Maxwell equation can be
expressed as $\nabla\times\delta{\bf B}=(4\pi/c)\delta {\bf J}$,
whose $z$-component is
\begin{eqnarray}
\left[\frac{1}{r}\frac{\partial}{\partial r}
\left(r\frac{\partial}{\partial r}\right)
+\frac{1}{r^{2}}\frac{\partial^{2}}{\partial\theta^{2}}
-\frac{4\pi\sigma(r)}{c^{2}}\frac{\partial}{\partial t}
\right]\delta A_{z}({\bf r},t)
\nonumber \\
=\frac{4\pi e}{c}\Delta f_{\rm pair}
\int{\rm d}^{3}{\bf p}v_{\rm b}\delta f_{\rm b}({\bf p},{\bf r},t),
\label{eq:25}
\end{eqnarray}
\noindent
for the cylindrically symmetric case.
Here, the terms proportional to
$c^{-2}(\partial^{2}/\partial t^{2})\delta A_{z}$
and $(\partial^{2}/\partial z^{2})\delta A_{z}$
have been neglected as consistent with the above assumptions.
In addition, the linearized Vlasov equation is described as
\begin{eqnarray}
\left[\frac{\partial}{\partial t}+v_{\rm b}\frac{\partial}{\partial z}
+e\beta_{\rm b}B_{\theta}^{\rm s}(r)\frac{\partial}{\partial p_{r}}\right]
\delta f_{\rm b}({\bf p},{\bf r},t)
\nonumber \\
=e\beta_{\rm b}\left[\frac{1}{r}
\frac{\partial\delta A_{z}({\bf r},t)}{\partial\theta}
\frac{\partial f_{\rm b}^{0}({\bf p},r)}{\partial p_{\theta}}
\right. \nonumber \\ \left.
+\frac{\partial\delta A_{z}({\bf r},t)}{\partial r}
\frac{\partial f_{\rm b}^{0}({\bf p},r)}{\partial p_{r}}
\right].
\label{eq:26}
\end{eqnarray}

In order to derive the dispersion relation from above set of
equations~(\ref{eq:25}) and (\ref{eq:26}), we assume the
standard Fourier expansions of the perturbed components:
\begin{equation}
\delta f_{\rm b}({\bf p},{\bf r},t)
=\delta f_{\rm b}({\bf p},r)
\exp\left[{\rm i}\left(\ell\theta +k_{z}z-\omega t\right)\right],
\label{eq:27}
\end{equation}
\begin{equation}
\delta A_{z}({\bf r},t)=\delta A_{z}(r)
\exp\left[{\rm i}\left(\ell\theta+k_{z}z-\omega t\right)\right],
\label{eq:28}
\end{equation}
\noindent
where ${\rm i}=\sqrt{-1}$ is the imaginary unit.
In equations~(\ref{eq:27}) and (\ref{eq:28}), it is convenient
to transform the set of the independent variables ($t$, $z$)
to ($\tau=t-z/v_{\rm b}$, $z$), so that
\begin{equation}
\delta f_{\rm b}({\bf p},r,\theta,z,\tau)
=\delta f_{\rm b}({\bf p},r)
\exp\left[{\rm i}\left(\ell\theta-\Omega z/v_{\rm b}
-\omega\tau\right)\right],
\label{eq:29}
\end{equation}
\begin{equation}
\delta A_{z}(r,\theta,z,\tau)=
\delta A_{z}(r)\exp\left[{\rm i}\left(\ell\theta
-\Omega z/v_{\rm b}-\omega\tau\right)\right],
\label{eq:30}
\end{equation}
where $\Omega=\omega-k_{z}v_{\rm b}$ is the frequency of the
perturbations seen by a beam electron \citep{davidson90}.
Below, we concentrate on an interesting case imposing the kink-type
perturbation with $\ell =1$ and seek the corresponding eigenmodes.

\section{MODIFIED EIGENVALUE EQUATIONS FOR KINK
PERTURBATION ON AN ELECTRON--POSITRON BEAM}

Using the analysis by \citet{uhm80}, we derive eigenvalue equations
from the self-consistent equations~(\ref{eq:25}) and (\ref{eq:26}):
(i) for the case of the rectangular density profile (Section~5.1)
and (ii) for the Bennett profile case (Section~5.2).

\subsection{The rectangular radial density profile case}

According to the derivations outlined in Appendix,
we derive an eigenvalue equation for the kink-type
perturbations superimposed on the filamentary flow.
For the rectangular density profile given in equation~(\ref{eq:8}),
we have $\partial n_{\rm b}/\partial r=-{\hat n}_{\rm b}\delta(r-r_{\rm b})$.
Then, the master equation~(\ref{eq:a9}) provides
the following eigenvalue equation:
\begin{eqnarray}
\left(\frac{1}{r}\frac{\partial}{\partial r}r\frac{\partial}{\partial r}
-\frac{1}{r^{2}}+\frac{4\pi{\rm i}\omega\sigma}{c^{2}}
\right)\delta A_{z}(r)
\nonumber \\
=\frac{\delta A_{z}(r){\hat\omega}_{\rm pb}^{2}\beta_{\rm b}^{2}
/\gamma_{\rm b}}
{\left(\Omega^{2}-{\hat\omega}_{\beta}^{2}\right)r}\delta(r-r_{\rm b}),
\label{eq:31}
\end{eqnarray}
\noindent
where the definition of
${\hat\omega}_{\rm pb}^{2}\equiv 4\pi e^{2}{\hat n}_{\rm b}
\Delta f_{\rm pair}/m_{\rm e}={\rm const}$ has been introduced.
It is evident, from the right-hand side (RHS) of equation~(\ref{eq:31}),
that the perturbed axial current is localized at the beam surface
($r=r_{\rm b}$), analogous to the Kelvin-Helmholtz-type instability
for neutral flows with local velocity shear \citep[e.g.][]{chandrasekhar81}.
Apparently, equation~(\ref{eq:31}) has the same form as
equation~(9.256) given in \citet{davidson90}, but now the
pair production rate is included in ${\hat\omega}_{\rm pb}$
and ${\hat\omega}_{\beta}$.
It is expected that in astronomical environments with
large pair production rates, both ${\hat\omega}_{\rm pb}$
and ${\hat\omega}_{\beta}$ are significantly reduced,
to lower the intensity of the perturbed surface current
and resonant frequency of $\Omega$, respectively.
When we impose the boundary condition of
$\delta A_{z}(r=0)=0=\delta A_{z}(r=\infty)$
and continuity of $\delta A_{z}$ at $r=r_{\rm b}$,
the eigenvalue contained in equation~(\ref{eq:31}) is
found to conform to the following dispersion relation:
\begin{equation}
\omega\tau_{\rm d}
={\rm i}\left[\frac{1}{\left(1-f_{\rm m}\right)\left(1-x\right)}-1\right].
\label{eq:32}
\end{equation}
Here, we introduce the definition of the
characteristic resistive decay time
\begin{equation}
\tau_{\rm d}\equiv\frac{\pi{\hat\sigma}r_{\rm b}^{2}}{2c^{2}},
\label{eq:33}
\end{equation}
\noindent
and $x\equiv\Omega^{2}/{\hat\omega}_{\beta}^{2}$,
and also use the relation of
${\hat\omega}_{\rm pb}^{2}\beta_{\rm b}^{2}
/(2\omega_{\beta}^{2}\gamma_{\rm b})=(1-f_{\rm m})^{-1}$,
where $f_{\rm m}\equiv(1-f_{\rm b})|\beta_{r}|/(f_{\rm b}\beta_{\rm b})$,
which takes a value in the range of $0\leq f_{\rm m}<1$
(cf. equation~\ref{eq:3}).
It is noted that equation~(\ref{eq:32}) is correct to
leading order in $\omega\tau_{\rm d}$.

\subsubsection{Mode property for real $\omega$ and complex $\Omega$}

First, we consider the real $\omega$ and complex $\Omega$ case;
the spatially growing/decaying mode may, for convenience,
be referred to as the convective mode.
To avoid confusion, it is noted that the growth of
perturbations can be viewed in the beam frame
(not in the wave frame as conventionally chosen).
In an interesting regime of $\omega\tau_{\rm d}\rightarrow 0$
(i.e. for low frequency and/or very high resistivity), the
left-hand side (LHS) of equation~(\ref{eq:32}) vanishes,
to yield the simple equation of $(1-f_{\rm m})(1-x)=1$,
which contains the solution of $x=-f_{\rm m}/(1-f_{\rm m})$.
Therefore, for $0<f_{\rm m}<1$, the value of
$x=k_{z}^{2}v_{\rm b}^{2}/{\hat\omega}_{\beta}^{2}$
is real and negative.
It follows that $k_{z}$ takes purely imaginary values so that
$k_{z}=\pm {\rm i}k_{\rm i}$, reflecting the purely growing
$-{\rm i}k_{\rm i}$ and purely decaying $+{\rm i}k_{\rm i}$
modes (cf. equations~\ref{eq:27} and \ref{eq:28}).
When the current neutral condition is perfectly satisfied
[i.e. $\beta=0$ ($f_{\rm m}=1$)], the value of
$k_{\rm i}=({\hat\omega}_{\beta}/v_{\rm b})[f_{\rm m}/(1-f_{\rm m})]^{1/2}$
diverges (and thus violates the theory), whereas no return
current ($f_{\rm m}=0$) trivially leads to $k_{\rm i}=0$.
Clearly, it turns out that the instability mechanism
in this regime is a result of the repulsion of the beam
current by the return current with $f_{\rm m}\neq 0$
\citep[for the $\omega\tau_{\rm d}\neq 0$ case, see][]{uhm80}.

We define the growth distance by $L=2\pi/k_{\rm i}$,
which scales as
\begin{eqnarray}
L&=&0.1
\left(\frac{\beta_{\rm b}/|\beta_{\rm r}|}{10}\right)^{1/2}
\left(\frac{\gamma_{\rm b}}{10}\right)^{1/2}
\nonumber \\
&\times&\left(\frac{10^{-3}}{\Delta f_{\rm pair}}\right)^{1/2}
\left(\frac{10^{-6}~{\rm cm^{-3}}}
{{\hat n}_{\rm r}}\right)^{1/2}~{\rm au}.
\label{eq:34}
\end{eqnarray}
\noindent
Here equation~(\ref{eq:12}), the relation of
$f_{\rm m}/(1-f_{\rm m})=(f_{\rm b}\beta_{\rm b}-\beta)/\beta
=(1-f_{\rm b})|\beta_{\rm r}|/\beta$ and
${\hat n}_{\rm r}=(1-f_{\rm b}){\hat n}_{\rm e}$ are used.
Importantly, as the pair production rate increases,
the value of $L$ increases, because of the apparent lowering
of ${\hat\omega}_{\beta}$ (cf. equation~\ref{eq:12}).
Even though the pair production effects increase $L$, the
predicted $L$ is much shorter than the length-scale of
major astrophysical jets, particularly, extragalactic jets.
Thus, the convectively growing mode (for lower $\omega$ and/or
shorter $\tau_{\rm d}$) seems not actually to reflect the
characteristic of the extremely stable transport of the jets.

\subsubsection{Mode property for complex $\omega$ and real $\Omega$}

Next we examine the complex $\omega$ and real $\Omega$ case;
the temporally growing/decaying mode may be referred to as
the absolute mode, using conventional terminology.
Because of real $\Omega$, $x$ $(=\Omega^{2}/{\hat\omega}_{\beta}^{2})$
is real and positive, so that the square bracket on the RHS
of equation~(\ref{eq:32}) is also real.
Thereby, $\omega$ is found to be purely imaginary:
$\omega={\rm i}\omega_{\rm i}$.
The complex wavelength is expressed in the form of
$k_{z}={\rm i}k_{\rm i}+k_{\rm r}$, and the relation
of $\Omega=\omega-k_{z}v_{\rm b}$ is recalled.
Then, we find $k_{\rm i}=\omega_{\rm i}/v_{\rm b}$
and $k_{\rm r}=-\Omega/v_{\rm b}$, which can,
respectively, be expressed as
\begin{equation}
k_{\rm i}=\frac{1}{v_{\rm b}\tau_{\rm d}}
\left[\frac{1}{\left(1-f_{\rm m}\right)\left(1-x\right)}-1\right],
\label{eq:35}
\end{equation}
\noindent
and
\begin{equation}
k_{\rm r}=-{\rm sgn}(\Omega)\frac{{\hat\omega}_{\beta}}{v_{\rm b}}\sqrt{x},
\label{eq:36}
\end{equation}
Here, ${\rm sgn}(\Omega\grole 0)=\pm 1$.
Note that the positive value of $k_{\rm i}$ can now be compared to
the growth factor for the absolute mode, and when $x$ approaches
unity below (i.e. $x\rightarrow 1-$), $k_{\rm i}\rightarrow +\infty$.
We can confirm that the destabilization involving
$k_{\rm i}>0$ takes place for $0<x<1$.
The fact of the stable transport of astrophysical jets
over parsec scales seems to be inconsistent with the
above prediction of the resonant destabilization.
As shown later in Section~5.2.2, the resonance can be
removed for the Bennett-type profile case.

\subsection{The modified Bennett profile case}

We again recall equation~(\ref{eq:a9}), multiplying through by
$r\delta A_{z}$ and integrating over $r$ from $0$ to $\infty$.
An appropriate choice of the trial function of
$\delta A_{z}^{\rm t}\propto\partial A_{z}^{\rm s}/\partial r$
corresponding to a rigid displacement of the magnetic
field \citep[cf.][for a complete discussion]{uhm80}
yields the extended eigenvalue equation of the form of
\begin{eqnarray}
\frac{{\rm i}\omega}{c}\int_{0}^{\infty}{\rm d}r~r\sigma(r)
\left[\frac{\partial A_{z}^{\rm s}(r)}{\partial r}\right]^{2}
=\int_{0}^{\infty}{\rm d}r~r\frac{\partial A_{z}^{\rm s}(r)}{\partial r}
\frac{\partial J(r)}{\partial r}
\nonumber \\
\times\left\{1+\frac{\beta_{\rm b}f_{\rm b}}{\beta}
\frac{\omega_{\beta}^{2}(r)}
{\left[\Omega^{2}-\omega_{\beta}^{2}(r)\right]}
\right\}.
\label{eq:37}
\end{eqnarray}
\noindent
Within the present framework (Section~2), the total current,
$J$ ($=J_{\rm b}+J_{\bar{\rm b}}+J_{\rm r}+J_{\bar{\rm r}}$),
can be expressed as
\begin{equation}
J(r)=\frac{\beta\Delta f_{\rm pair}}
{\beta_{\rm b}f_{\rm b}}J_{\rm b}(r),
\label{eq:38}
\end{equation}
\noindent
where $J_{\rm b}(r)=ef_{\rm b}n_{\rm e}(r)v_{\rm b}$.
Using equations~(\ref{eq:15}), (\ref{eq:17}), (\ref{eq:22})
and (\ref{eq:38}), and the variable exchange of
$\xi=(1+r^{2}/r_{\rm b}^{2})^{-1}$,
equation~(\ref{eq:37}) is cast to
\begin{eqnarray}
\frac{{\rm i}\pi r_{\rm b}^{2}\omega}{2c^{2}}
\int_{0}^{1}{\rm d}\xi\frac{1-\xi}{\xi}\sigma(\xi)
=\int_{0}^{1}{\rm d}\xi\xi\left(1-\xi\right)
\nonumber \\
\times\frac{\Omega^{2}+\left[\xi{\hat\omega}_{\beta}^{2}
\left(1-f_{\rm b}\right)|\beta_{\rm r}|/\beta\right]}
{\Omega^{2}-\xi{\hat\omega}_{\beta}^{2}}.
\label{eq:39}
\end{eqnarray}
\noindent
Note that $r_{\rm b}^{2}$ and ${\hat\omega}_{\beta}$ involve the
pair production rate, although the form of equation~(\ref{eq:39})
itself is similar to equation~(73) given in \citet{uhm80}.
The RHS of equation~(\ref{eq:39}) can be exactly
integrated, to give
\begin{equation}
\omega\tau_{\rm d}
={\rm i}\left(1-f_{\rm m}\right)^{-1}\left[G(x; g)+f_{\rm m}\right].
\label{eq:40}
\end{equation}
\noindent
Here, we introduce the definitions of
\begin{equation}
\tau_{\rm d}\equiv\frac{3\pi r_{\rm b}^{2}}{c^{2}}
\int_{0}^{1}{\rm d}\xi\frac{1-\xi}{\xi}\sigma(\xi),
\label{eq:41}
\end{equation}
\noindent
and
\begin{equation}
G(x; g)\equiv 6x\left\{\frac{1}{2}-x+x\left(1-x\right)
\left[\ln\left|\frac{1-x}{x}\right|+{\rm i}g\pi\right]
\right\},
\label{eq:42}
\end{equation}
\noindent
where $g=0$ for $x<0$ and $x>1$; $g={\rm sgn}(\Omega)$ for $0<x<1$.
Note that when setting the expression of the electrical
conductivity to $\sigma(\xi)={\hat\sigma}\xi^{2}$,
where ${\hat\sigma}={\rm const}$, equation~(\ref{eq:41})
coincides with equation~(\ref{eq:33}).

\subsubsection{Mode property for real $\omega$ and complex $\Omega$}

In parallel to the argument given in Section~5.1.1,
we examine the real $\omega$ and complex $\Omega$ case,
for convenience, referred to as the convective mode.
In an interesting regime of $\omega\tau_{\rm d}\rightarrow 0$,
the square bracket of the RHS of equation~(\ref{eq:40}) vanishes,
to give the dispersion relation of $G(x; g)+f_{\rm m}=0$.
In particular, the equation of $G(x; g=0)+f_{\rm m}=0$
is found to contain the solution of $x=\chi(f_{\rm m})$
that is real and, for the possible range of $f_{\rm m}$,
negative (as consistent with the choice of $g=0$).
Hence, recalling the expression of
$x=k_{z}^{2}v_{\rm b}^{2}/{\hat\omega}_{\beta}^{2}$,
the dispersion relation turns out to contain the complex $k_{z}$,
which is purely imaginary so that $k_{z}=\pm{\rm i}k_{\rm i}$,
reflecting the purely growing $-{\rm i}k_{\rm i}$ and
purely decaying $+{\rm i}k_{\rm i}$ mode.
Note here that
$k_{\rm i}=({\hat\omega}_{\beta}/v_{\rm b})[-\chi(f_{\rm m})]^{1/2}$.

\begin{figure}
\includegraphics[width=100mm]{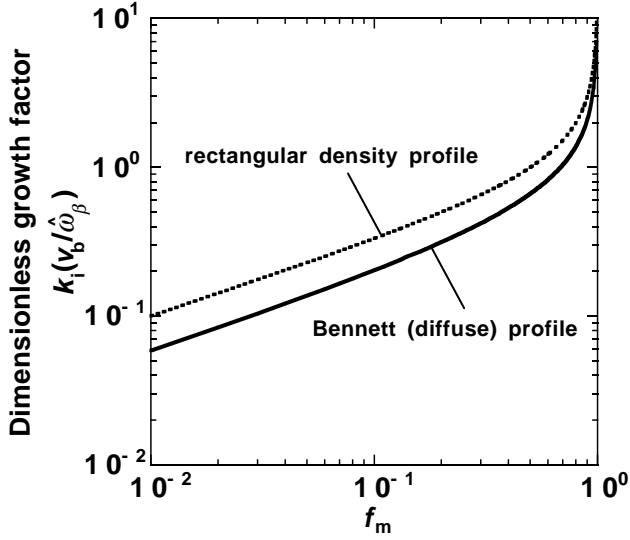}
\caption{The growth factor of a convective mode,
$k_{\rm i}(\omega\tau_{\rm d}\rightarrow 0)$, multiplied by
$v_{\rm b}/{\hat\omega}_{\beta}$ as a function of $f_{\rm m}$,
for the rectangular radial density profile (equation~\ref{eq:8};
dotted curve) and the diffuse radial density profile
for the modified Bennett pinch equilibrium
(equation~\ref{eq:15}; solid curve).
Here, $\tau_{\rm d}$ is the resistive decay time
(equations~\ref{eq:33} and \ref{eq:41}), $v_{\rm b}$ is
the beam velocity, ${\hat\omega}_{\beta}$ is the betatron
frequency at the beam centre $r=0$ (equation~\ref{eq:12}), and
$f_{\rm m}$ is the ratio of the return/beam current density.}
\label{f1}
\end{figure}

Fig.~\ref{f1} plots the dimensionless growth factor
$(-\chi)^{1/2}$ as a function of $f_{\rm m}$ for $0.01\leq f_{\rm m}<1$,
together with, for comparison, the function of
$k_{\rm i}(v_{\rm b}/{\hat\omega}_{\beta})=[f_{\rm m}/(1-f_{\rm m})]^{1/2}$
for the case of the rectangular radial density profile (Section~5.1.1).
When the current neutral condition is perfectly satisfied
[i.e. $\beta=0$ ($f_{\rm m}=1$)], the value of
$k_{\rm i}=({\hat\omega}_{\beta}/v_{\rm b})[-\chi(f_{\rm m})]^{1/2}$
diverges, whereas no return current ($f_{\rm m}=0$)
trivially leads to $k_{\rm i}=0$.
It is, again, found that the repulsive interaction
between the beam current and the return current is
essential for driving the instability.
As seen in Fig.~\ref{f1}, the present value of $k_{\rm i}$ is
smaller than that obtained for the rectangular density profile.
Namely, the diffuse boundary effects extend
the growth distance of $L=2\pi/k_{\rm i}$.
However, the effects are not so significant that the expected
value of $L$ is found to be much shorter than the length-scale
of astrophysical jets (cf. equation~\ref{eq:34}).
More importantly, the pair production effects significantly
lower the value of ${\hat\omega}_{\beta}$, thereby decreasing
$k_{\rm i}$; however, the value of $L$ seems to be still
much shorter than the observed length of the jets.

\subsubsection{Mode property for complex $\omega$ and real $\Omega$}

In parallel to Section~5.1.2, we examine the complex $\omega$
and real $\Omega$ case, referred to as the absolute mode.
Because of real $\Omega$, $x$ $(=\Omega^{2}/{\hat\omega}_{\beta}^{2})$
is real and positive, so that $G(x; g=0)$ is also real.
For $g\ne 0$ (for $0<x<1$), both $k_{z}$ and $\omega$
are complex, which may be expressed as $k_{z}={\rm i}k_{\rm i}+k_{\rm r}$
and $\omega={\rm i}\omega_{\rm i}+\omega_{\rm r}$.
Recalling equations~(\ref{eq:40}) and (\ref{eq:42}),
and the relation $\Omega=\omega-k_{z}v_{\rm b}$,
we have $k_{\rm i}=\omega_{\rm i}/v_{\rm b}$ and
$k_{\rm r}=(\omega_{\rm r}-\Omega)/v_{\rm b}$,
which can, respectively, be expressed as
\begin{equation}
k_{\rm i}=\frac{1}{v_{\rm b}\tau_{\rm d}}
\frac{G_{0}(x)+f_{\rm m}}{1-f_{\rm m}},
\label{eq:43}
\end{equation}
\noindent
where $G_{0}(x)\equiv G(x; g=0)$, and
\begin{equation}
k_{\rm r}=-\frac{1}{v_{\rm b}}
\left[g\pi\frac{6x^{2}\left(1-x\right)}
{\left(1-f_{\rm m}\right)\tau_{\rm d}}
+{\rm sgn}(\Omega){\hat\omega}_{\beta}\sqrt{x}
\right].
\label{eq:44}
\end{equation}
\noindent
On the RHS of equation~(\ref{eq:44}), the first term
inside the square bracket is ordinarily much smaller
than the second term, whereupon equation~(\ref{eq:44})
can be approximated by equation~(\ref{eq:36}).
The positive value of $k_{\rm i}$ can be compared to
the growth factor for the absolute mode, but,
importantly, equation~(\ref{eq:43}) does not contain
the resonance point of $x$ that appeared in the
rectangular density profile (Section~5.1.2).

\begin{figure}
\includegraphics[width=90mm]{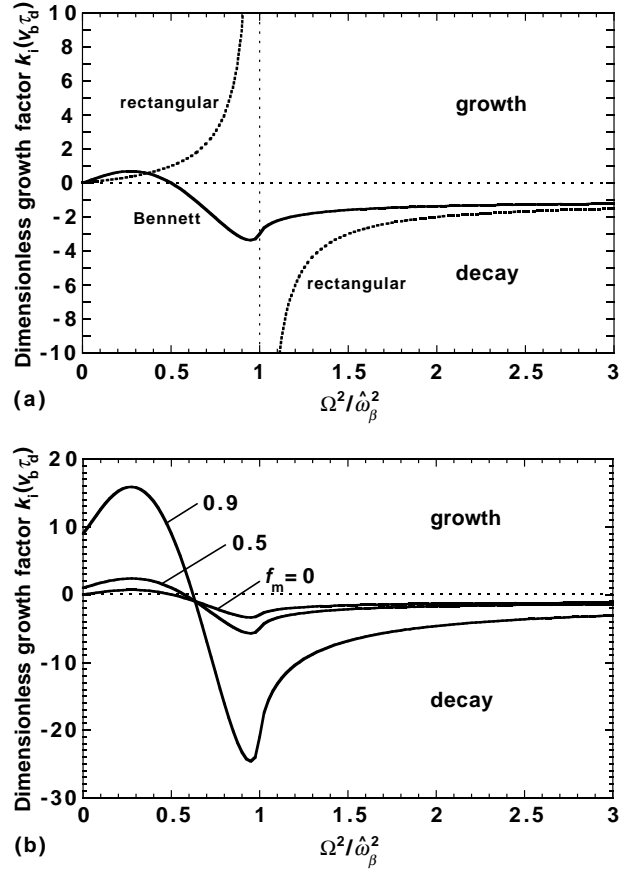}
\caption{The growth factor of the absolute mode, $k_{\rm i}$,
multiplied by $v_{\rm b}\tau_{\rm d}$ as a function of
$\Omega^{2}/{\hat\omega}_{\beta}^{2}$, indicating
(a) for $f_{\rm m}=0$, a comparison in between the rectangular
profile (equation~\ref{eq:35}; dotted curve) and modified Bennett
profile (equation~\ref{eq:43}; solid curve) cases, and
(b) the $f_{\rm m}$-dependence for the latter case.
Here, $\Omega$ $(\equiv\omega-k_{z}v_{\rm b})$ is the
frequency of perturbations viewed in the beam frame.
In (a), for the rectangular case the resonant point
exists at $\Omega^{2}/{\hat\omega}_{\beta}^{2}=1$
(vertical thin dashed line).
Note that the solid curve in (a) and the curve for $f_{\rm m}=0$ in (b)
correspond to the function $G(\Omega^{2}/{\hat\omega}_{\beta}^{2}; g=0)$
(equation~\ref{eq:42}; \citealt{uhm80}).}
\label{f2}
\end{figure}

Fig.~\ref{f2}(a) plots, for the special case of $f_{\rm m}=0$,
the dimensionless growth factor $k_{\rm i}(v_{\rm b}\tau_{\rm d})$
of equation~(\ref{eq:43}) as a function of $x$ in the range
$x\geq 0$, together with, for comparison, equation~(\ref{eq:35})
for the rectangular profile case.
It is noted that for $f_{\rm m}=0$, the function
$k_{\rm i}(v_{\rm b}\tau_{\rm d})$ of equation~(\ref{eq:43})
reduces to $G_{0}(x)$, recalling fig.~2 in \citet{uhm80}
and fig.~9.16 in \citet{davidson90}.
As $G_{0}(x\rightarrow 0)=0$ and $G_{0}(x\rightarrow +\infty)=-1$,
in equation~(\ref{eq:43}) we find
$k_{\rm i}(x\rightarrow 0)=(v_{\rm b}\tau_{\rm d})^{-1}
[f_{\rm m}/(1-f_{\rm m})]$ and
$k_{\rm i}(x\rightarrow +\infty)=-(v_{\rm b}\tau_{\rm d})^{-1}$;
these asymptotic values coincide with those of equation~(\ref{eq:35}).
As seen in Fig.~\ref{f2}(a), in the range $0<x<1$,
the growth factor for the rectangular case is always positive,
whereas the diffuse effects arrange an $x$-domain in which
the stable condition, $k_{\rm i}\leq 0$, is satisfied.
Note that $G_{0}(x=\frac{1}{2})=0$; in the range
$0<x<\frac{1}{2}$, $G_{0}(x)>0$ takes the maximum
value of $G_{0,\rm max}=0.690$ at $x=0.272$, whereas
in the range $x>\frac{1}{2}$, $G_{0}(x)<0$ takes the
minimum value of $G_{0,\rm min}=-3.36$ at $x=0.949$.
In particular, for $x>0.635$, $G_{0}(x)<-1$, so that
the stable condition, $k_{\rm i}<0$, is always
satisfied for the possible range of $f_{\rm m}$.

Fig.~\ref{f2}(b) plots, for $f_{\rm m}=0$, $0.5$
and $0.9$, $k_{\rm i}(v_{\rm b}\tau_{\rm d})$
(equation~\ref{eq:43}) as a function of $x$.
It is clearly found that when the return current increases
(i.e. with increasing $f_{\rm m}$), the maximum (minimum)
value of $k_{\rm i}(v_{\rm b}\tau_{\rm d})$, which is achieved
at the common $x$-point of $x|_{G_{0}=G_{0,{\rm max}}}$
($x|_{G_{0}=G_{0,{\rm min}}}$), enormously increases (decreases).
In contrast to the infinite divergence at $x=1$ involved
in equation~(\ref{eq:35}), $G_{0,\rm max}$ is found to
provide the finite maximum value of $k_{\rm i}$
[i.e. $k_{\rm i,max}\simeq(v_{\rm b}\tau_{\rm d})^{-1}
(0.7+f_{\rm m})(1-f_{\rm m})^{-1}$ at $x\simeq 0.27$],
where equation~(\ref{eq:44}) approximately reads
$k_{\rm r}\simeq -{\rm sgn}(\Omega)0.52{\hat\omega}_{\beta}/v_{\rm b}$.
Namely, the growth factor is bounded; physically, this is
because there is no single frequency $\Omega$ for which
the entire beam is resonant with the wave \citep{uhm80}.
The values of $k_{\rm i,max}$ and $k_{\rm r}$ determine
the shortest growth distance of wave envelope defined
by $L_{\rm min}\equiv 2\pi/k_{\rm i,max}$ and the
corresponding oscillation wavelength of the carrier wave,
$\lambda_{\rm osc}\equiv 2\pi/|k_{\rm r}|$, respectively.
These are explicitly written as
\begin{equation}
L_{\rm min}=\frac{\pi^{2}\beta_{\rm b}r_{\rm b}^{2}{\hat\sigma}}{c}
\frac{1-f_{\rm m}}{0.7+f_{\rm m}},
\label{eq:45}
\end{equation}
\begin{equation}
\lambda_{\rm osc}=\frac{12v_{\rm b}}{{\hat\omega}_{\beta}}.
\label{eq:46}
\end{equation}
In the derivation of equation~(\ref{eq:45}), equation~(\ref{eq:33})
has been employed as the expression of $\tau_{\rm d}$.
The obtained result suggests that astrophysical jets
can propagate over, at least, the distance of
$\sim L_{\rm min}$, having the spatial oscillation with
the wavelength of $\lambda_{\rm osc}$ $(\ll L_{\rm min})$.

Let us evaluate equation~(\ref{eq:45}).
Concerning the electrical conductivity, for simplicity
we here employ the standard Spitzer formula.
In the regime of $T\gg m_{\rm e}c^{2}$ likely
for pair plasmas, we have the relation
${\hat\sigma}/\sigma_{\rm S}\sim[(2\pi)^{1/2}/8](m_{\rm e}c^{2}/T)^{1/2}$,
where $\sigma_{\rm S}\sim T^{3/2}/(e^{2}m_{\rm e}^{1/2}\ln\Lambda)$
is the well-known non-relativistic Spitzer conductivity
\citep{braams89,honda03}; that is, for the Coulomb logarithm of
$\ln\Lambda\sim 10$ \citep[e.g.][]{huba94}, the scaling of
${\hat\sigma}\sim 10^{12}(T/10^{10}~{\rm K})~{\rm s^{-1}}$.
We also call the relation
$(1-f_{\rm m})/(0.7+f_{\rm m})
=\beta/(1.7f_{\rm b}\beta_{\rm b}-\beta)$.
Then, using equations~(\ref{eq:16}) and the above
expression of ${\hat\sigma}$ for $T\gg m_{\rm e}c^{2}$,
equation~(\ref{eq:45}) is found to scale as
\begin{equation}
L_{\rm min}=2.7\eta
\left(\frac{10}{\ln\Lambda}\right)
\left(\frac{1~{\rm cm^{-3}}}{{\hat n}_{\rm b}}\right)
\left(\frac{T}{10^{10}~{\rm K}}\right)^{2}~{\rm Mpc},
\label{eq:47}
\end{equation}
while for $T\ll m_{\rm e}c^{2}$, setting to
${\hat\sigma}=\sigma_{\rm S}$, yields
\begin{equation}
L_{\rm min}=1.1\times 10^{2}\eta
\left(\frac{10}{\ln\Lambda}\right)
\left(\frac{1~{\rm cm^{-3}}}{{\hat n}_{\rm b}}\right)
\left(\frac{T}{10^{8}~{\rm K}}\right)^{5/2}~{\rm pc}.
\label{eq:48}
\end{equation}
Here, the enhancement factor is introduced, defined as
\begin{equation}
\eta\equiv\frac{1-\left(\Delta f_{\rm pair}/2\right)}
{\beta\Delta f_{\rm pair}^{2}
\left[1-\left(\beta/1.7f_{\rm b}\beta_{\rm b}\right)\right]},
\label{eq:49}
\end{equation}
\noindent
which takes a value in the range
$\eta\geq 2/(1.7f_{\rm b}\beta_{\rm b})$.
The strong $T$-dependence of equations~(\ref{eq:47}) and (\ref{eq:48})
arises from that of $r_{\rm b}^{2}\propto T$, and ${\hat\sigma}\propto T$
(for $T\gg m_{\rm e}c^{2}$) or $T^{3/2}$ (for $T\ll m_{\rm e}c^{2}$).

In the regimes of $\Delta f_{\rm pair}/2\ll 1$
and $\beta/(1.7f_{\rm b}\beta_{\rm b})\ll 1$,
equation~(\ref{eq:49}) can be approximated by
$\eta\simeq 1/(\beta\Delta f_{\rm pair}^{2})\gg 1$.
In particular, for $\Delta f_{\rm pair}/2\ll 1$, the factor $\eta$
is reciprocal to $\Delta f_{\rm pair}^{2}$ (i.e. proportional to
the square of the pair production rate), reflecting that the
radial expansion stemming from the charge screening effects by
positrons ($r_{\rm b}\propto\Delta f_{\rm pair}^{-1}$) significantly
increases the value of $\tau_{\rm d}$, to extend $L_{\rm min}$.
For $\beta<0.85f_{\rm b}\beta_{\rm b}$, $\eta$ increases
with decreasing $\beta$, and particularly for
$\beta\ll 1.7f_{\rm b}\beta_{\rm b}$,
$\eta\propto\beta^{-1}$; this implies that the
radial expansion effect ($r_{\rm b}\propto\beta^{-1}$)
tends to surpass the destabilization by the return current.
Even for the special case of $f_{\rm b}=1$ and
$\beta=\beta_{\rm b}$ ($f_{\rm m}=0$),
equations~(\ref{eq:47}) and (\ref{eq:48}) are valid,
having $\eta=(2.4/\beta_{\rm b})[1-(\Delta f_{\rm pair}/2)]
/\Delta f_{\rm pair}^{2}$ ($\geq 1.2$).

\begin{figure}
\includegraphics[width=108mm]{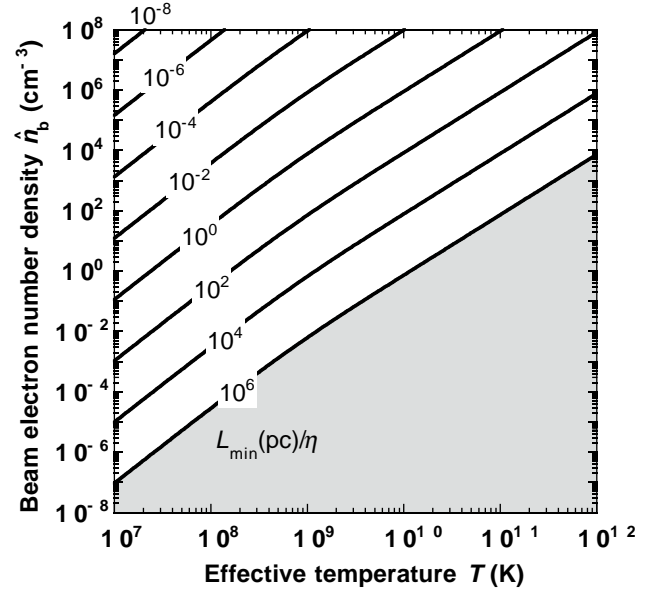}
\caption{The contour plots of $L_{\rm min}/\eta$ for the
values of $10^{-8}-10^{6}~{\rm pc}$ (labelled) in the
${\hat n}_{\rm b}-T$ parameter space, for the Bennett profile case
in Fig.~\ref{f2} (exhibiting the maxima $k_{\rm i,max}$).
Here, $L_{\rm min}$ is the shortest growth distance (defined by
$L_{\rm min}\equiv 2\pi/k_{\rm i,max}$; equation~\ref{eq:45}),
$\eta$ $(>1)$ is the enhancement factor (equation~\ref{eq:49}),
${\hat n}_{\rm b}$ is the beam electron number density at $r=0$,
and $T$ is the effective temperature (equation~\ref{eq:4}).
The shaded area indicates the (${\hat n}_{\rm b}$, $T$)-parameter
domain where supra-Mpc-scale transport is allowed for $\eta\sim 1$.}
\label{f3}
\end{figure}

Fig.~\ref{f3} plots ${\hat n}_{\rm b}$ as a function
of $T$, for the given parameter $L_{\rm min}/\eta$
in the range $10^{-8}-10^{6}~{\rm pc}$.
Here, the generic formulae of ${\hat\sigma}$ is used, which
covers the mild relativistic region \citep{braams89,honda03}
and $\ln\Lambda$ dependent weakly on density and temperature
\citep[e.g. for ${\hat n}_{\rm b}\geq 1~{\rm cm^{-3}}$
and $T\leq 10^{10}~{\rm K}$, $\ln\Lambda\la 30$;][]{huba94}.
The shaded area indicates the (${\hat n}_{\rm b}$, $T$)-parameter
domain in which Mpc-scale transport of jets is safely achieved
even for the marginal (most pessimistic) case of $\eta\sim 1$.
Such a parameter domain more significantly expands for the larger
values of $\eta$ owing to, for example, the larger pair production rate.
As a consequence, it can be claimed that there is a wide
parameter domain in which relativistic outflow can
safely propagate up to $1~{\rm Mpc}$ corresponding to
the observed longest scale of extragalactic jets.
Also, for fixed $\eta$, the smaller $L_{\rm min}$ allows the flow
to possess lower $T$ and higher ${\hat n}_{\rm b}$, that is,
expanding the allowable (${\hat n}_{\rm b}$, $T$)-parameter domain.
This property seems to be reflected in small-scale jets
such as Galactic jets, comprising lower temperature,
electron--ion plasma with $T\ll m_{\rm e}c^{2}$ and
$f_{\rm pair}=0$ ($\Delta f_{\rm pair}=1$).

\section{DISCUSSION AND CONCLUSIONS}

As investigated above, absolute growth can be allowed,
particularly for the case of the diffuse radial density
profile reflecting the Bennett pinch-type equilibrium.
However, in a low-frequency and/or high-resistivity regime,
the characteristic distance of the convective growth is,
for both the rectangular density and Bennett profiles,
much shorter than the observed length of astrophysical
jets, even if the pair production rate is large enough
to lower the betatron frequency ${\hat\omega}_{\beta}$.
Hence, the actual development of the convective mode
appears to be ruled out.
A simple explanation for the reason why the mode could
not crucially evolve is that the condition for zero
return current, namely $f_{\rm m}=0$, is perfectly
satisfied inside the beam \citep{honda00a}.
Also, if the mean poloidal (axial) magnetic field, $B_{0}{\bf\hat z}$,
is superimposed to arrange the helical field \citep[e.g.][]
{asada02,gabuzda03}, as well as the plasma wall in the
vacuum channel (Section~1) being in close proximity to the
beam, the mode can be stabilized \citep{uhm80,davidson90}.
For example, for the rectangular profile case, the sufficient
condition for instability can be written as
\begin{equation}
f_{\rm m}>\frac{\left(r_{\rm b}/b\right)^{2}+\varrho^{2}}{1+\varrho^{2}},
\label{eq:50}
\end{equation}
\noindent
where $b$ $(>r_{\rm b})$ is the radial distance from $r=0$ to the
plasma wall, $\varrho\equiv\omega_{\rm cb}/(2{\hat\omega}_{\beta})$,
and $\omega_{\rm cb}\equiv eB_{0}/(\gamma_{\rm b}m_{\rm e}c)$
is the relativistic cyclotron frequency.
It turns out, in equation~(\ref{eq:50}), that the aforementioned
cases of either $f_{\rm m}=0$, $\varrho^{2}\gg 1$ realized
for a larger $B_{0}$ (or a smaller $\gamma_{\rm b}$) or
$r_{\rm b}/b\sim 1$ lead to the robust stabilization.
Another possible explanation is that because the plasma is
likely non-uniform in actual environments, the mode can be
continuously carried away from the unstable (to stable) region.
However, the absolute growing mode tends to be retained at
the region where the unstable condition is satisfied, thereby
the instability can evolve (but note again that in the
present context, the growth factor will be favorably small,
as reflected in equations~\ref{eq:47} and \ref{eq:48}).
Note that a similar mode-preference is known to appear
in laboratory plasmas establishing non-uniformity.

Along with equation~(\ref{eq:50}), once the aforementioned
necessary steps are taken to stabilize the convective mode,
the stability properties of the absolute mode also change.
Similar to the convective mode, for example. for the rectangular
profile case, the larger values of $r_{\rm b}/b$ and
$f_{\rm m}=0$ have a stabilizing influence on the absolute mode.
The superimposition of $B_{0}\neq 0$ yields different
effects: shifting the range of $\Omega(=\omega-k_{z}v_{b})$
corresponding to instability relative to the
$B_{0}=0$ case, and expanding the bandwidth
$(\omega_{\rm cb}^{2}+{\hat\omega}_{\beta}^{2})^{1/2}$
of the instability in the $\Omega$-space \citep{davidson90}.
It appears that these effects are not significant enough
to disturb the reasoning that the convective mode must be
somehow strongly suppressed or stabilized in the actual system.

In conclusion, a full derivation has been given of the
eigenvalue equation for the kink-type perturbations
superimposed on the relativistic electron--positron beam
with the modified Bennett and rectangular profiles.
The dispersion relations have been examined for the convective
and absolute modes of the resistive hose instability.
It has been shown explicitly that the pair production
effects lower the betatron frequency ${\hat\omega}_{\beta}$
and expand the beam filament radius $r_{\rm b}$.
The following key results are found.
\begin{enumerate}
\item The reduction of ${\hat\omega}_{\beta}$ leads to a
reduction of the growth factor of a convective mode
scaled as $k_{\rm i}\sim{\hat\omega}_{\beta}/v_{\rm b}$.
\item The expansion of $r_{\rm b}$ prolongs the resistive decay time
for perturbed current $\tau_{\rm d}$ ($\propto r_{\rm b}^{2}$;
equation~\ref{eq:33}), resulting in a reduction of the growth factor of
the absolute mode scaled as $k_{\rm i}\sim 1/(v_{\rm b}\tau_{\rm d})$.
\end{enumerate}
For typical parameters, we found ${\hat\omega}_{\beta}\gg 1/\tau_{\rm d}$,
implying that the convective growth is dominant, although
its enormously rapid evolution, which ought to disrupt jets,
appears to be incompatible with the observational facts
because of the possible interpretation given above.

The Bennett profile effects also suppress the
convective growth, but the growth distance is still
much shorter than the observed length of jets
(even if the positron effects are involved).
In addition, the effects remove, for the absolute mode,
the resonant divergence of the growth factor, and
the growth distance is found to be comparable to
(or longer than) the length-scale of jets.
To summarize these consequences, preferentially,
(i) actual evolution of the convective mode, in a low
frequency and/or high resistivity regime, is ruled out
and (ii) evolution of the absolute mode is allowed,
more safely for the Bennett profile case.
Closer inspections by observations, laboratory
experiments and kinetic simulations are awaited.
\\

\appendix
\section{DERIVATION OF MODIFIED EIGENVALUE EQUATION}

In this appendix, we derive an eigenvalue equation that
governs equations~(\ref{eq:31}) and (\ref{eq:37}).
According to the method of characteristics \citep{krall86,davidson90},
we begin by integrating equation~(\ref{eq:26}) from
$z^{\prime}=-\infty$ to $z^{\prime}=z$.
Neglecting the initial perturbation at $z^{\prime}=-\infty$,
we obtain
\begin{eqnarray}
\delta f_{\rm b}({\bf p},r,\theta,z,\tau)=
\frac{e}{c}\frac{p_{z}}{p_{\perp}}
\frac{\partial f_{\rm b}^{0}({\bf p},r)}{\partial p_{\perp}}
\int_{-\infty}^{z}\frac{{\rm d}z^{\prime}}{v_{\rm b}}
\nonumber \\
\times\left(v_{r}^{\prime}\frac{\partial}{\partial r^{\prime}}
+\frac{v_{\theta}^{\prime}}{r^{\prime}}
\frac{\partial}{\partial\theta^{\prime}}
\right)
\delta A_{z}[r^{\prime}(z^{\prime}),
\theta^{\prime}(z^{\prime}),z^{\prime},\tau].
\label{eq:a1}
\end{eqnarray}
\noindent
Using
$v_{r}^{\prime}=v_{\rm b}\left({\rm d}r^{\prime}/{\rm d}z^{\prime}\right)$
and $v_{\theta}^{\prime}=
r^{\prime}v_{\rm b}\left({\rm d}\theta^{\prime}/{\rm d}z^{\prime}\right)$,
and equations~(\ref{eq:29}) and (\ref{eq:30}),
equation~(\ref{eq:a1}) can be written as
\begin{eqnarray}
\delta f_{\rm b}({\bf p},r)=\frac{e}{c}\frac{p_{z}}{p_{\perp}}
\frac{\partial f_{\rm b}^{0}({\bf p},r)}{\partial p_{\perp}}
\int_{-\infty}^{z}{\rm d}z^{\prime}
\left(
\frac{{\rm d}r^{\prime}}{{\rm d}z^{\prime}}
\frac{\partial}{\partial r^{\prime}}
+\frac{{\rm d}\theta^{\prime}}{{\rm d}z^{\prime}}
\frac{\partial}{\partial\theta^{\prime}}
\right)
\nonumber \\
\times\delta A_{z}[r^{\prime}(z^{\prime})]
\exp\left\{{\rm i}\left[\theta^{\prime}(z^{\prime})-\theta\right]
-{\rm i}\frac{\Omega}{v_{\rm b}}\left(z^{\prime}-z\right)\right\}.
\label{eq:a2}
\end{eqnarray}
\noindent
In terms of the generic function of
$\delta{\cal F}[r^{\prime}(z^{\prime}),
\theta^{\prime}(z^{\prime}),z^{\prime}]$,
we utilize the chain rule for differentiation:
${\rm d}\delta{\cal F}/{\rm d}z^{\prime}=\left[
\left({\rm d}r^{\prime}/{\rm d}z^{\prime}\right)
\left(\partial/\partial r^{\prime}\right)
+\left({\rm d}\theta^{\prime}/{\rm d}z^{\prime}\right)
\left(\partial/\partial\theta^{\prime}\right)
+\partial/\partial z^{\prime}\right]
\delta{\cal F}$.
Then, we have
\begin{eqnarray}
\delta f_{\rm b}({\bf p},r)=
\frac{e}{c}\frac{p_{z}}{p_{\perp}}
\frac{\partial f_{\rm b}^{0}({\bf p},r)}{\partial p_{\perp}}
\int_{-\infty}^{z}{\rm d}z^{\prime}\left(
\frac{\rm d}{{\rm d}z^{\prime}}+{\rm i}\frac{\Omega}{v_{\rm b}}
\right)
\nonumber \\
\times\delta A_{z}[r^{\prime}(z^{\prime})]
\exp\left\{{\rm i}\left[\theta^{\prime}(z^{\prime})-\theta\right]
-{\rm i}\frac{\Omega}{v_{\rm b}}\left(z^{\prime}-z\right)\right\}.
\label{eq:a3}
\end{eqnarray}
\noindent
When we impose $\theta^{\prime}(z^{\prime}=z)=\theta$
and $r^{\prime}(z^{\prime}=z)=r$, and recall
$\left[\delta A_{z}(r^{\prime})\right]_{z^{\prime}=-\infty}=0$,
the integration of equation~(\ref{eq:a3}) gives
\begin{eqnarray}
\delta f_{\rm b}({\bf p},r)=\frac{e}{c}\frac{p_{z}}{p_{\perp}}
\frac{\partial f_{\rm b}^{0}({\bf p},r)}{\partial p_{\perp}}
\left[\delta A_{z}(r)+{\rm i}\frac{\Omega}{v_{\rm b}}
\int_{-\infty}^{z}{\rm d}z^{\prime}
\right. \nonumber \\ \left.
\times\delta A_{z}[r^{\prime}(z^{\prime})]
\exp\left\{{\rm i}\left[\theta^{\prime}(z^{\prime})-\theta\right]
-{\rm i}\frac{\Omega}{v_{\rm b}}\left(z^{\prime}-z\right)\right\}\right].
\label{eq:a4}
\end{eqnarray}
\noindent
Substituting equation~(\ref{eq:a4}) into equation~(\ref{eq:25})
concomitant with the replacement of
$\partial/\partial t\rightarrow\partial/\partial\tau$,
we obtain
\begin{eqnarray}
\left[\frac{1}{r}\frac{\partial}{\partial r}
\left(r\frac{\partial}{\partial r}\right)
-\frac{1}{r^{2}}
+\frac{4\pi{\rm i}\omega\sigma(r)}{c^{2}}
\right]\delta A_{z}(r)
\nonumber \\
=4\pi e^{2}\beta_{\rm b}^{2}{\hat n}_{\rm b}\Delta f_{\rm pair}
\int_{0}^{\infty}{\rm d}U\frac{\partial}{\partial U}
F_{\rm b}(H)
\nonumber \\
\times\left[\delta A_{z}(r)+\frac{\Omega}{v_{\rm b}}I(\Omega,r,U)\right],
\label{eq:a5}
\end{eqnarray}
where the orbit integral $I$ is defined as
\begin{eqnarray}
I(\Omega,r,U)\equiv{\rm i}\int_{0}^{2\pi}\frac{{\rm d}\phi}{2\pi}
\int_{-\infty}^{0}{\rm d}\zeta
\delta A_{z}[r^{\prime}(z^{\prime})]
\nonumber \\
\times\exp\left\{{\rm i}\left[\theta^{\prime}(z^{\prime})-\theta\right]
-{\rm i}\frac{\Omega}{v_{\rm b}}\zeta\right\}.
\label{eq:a6}
\end{eqnarray}
\noindent
Here, $\zeta=z^{\prime}-z$, and $\phi$ is the
perpendicular momentum phase, conforming to
$p_{x}+{\rm i}p_{y}=p_{\perp}(\cos\phi+{\rm i}\sin\phi)$.

Concerning the unknown function $\delta A_{z}[r^{\prime}(z^{\prime})]$
in the integrand of equation~(\ref{eq:a6}), it is known that,
for $|\omega|\tau_{\rm d}<1$ (where $\tau_{\rm d}$ is the
resistive decay time; cf. equations~\ref{eq:33} and \ref{eq:41}),
$\delta A_{z}(r^{\prime})=\delta A_{z}(r)r^{\prime}/r$ provides a
reasonably good approximation in the beam interior permeated by the
magnetic field of $B_{\theta}^{\rm s}\propto r$ \citep{davidson90}.
Physically, such a solution reflects a
rigid displacement of the magnetic field.
Putting the approximate expression of $\delta A_{z}(r^{\prime})$
into equation~(\ref{eq:a6}), the betatron orbit expression
$r^{\prime}(z^{\prime})\exp\left[{\rm i}\theta^{\prime}(z^{\prime})\right]$
arises in the integrand.
As usual, this can be expanded by solving the equation
of motion of $\{v_{\rm b}({\rm d}^{2}/{\rm d}z^{\prime^{2}})
+\omega_{\beta}^{2}[r^{\prime}(z^{\prime})]\}
[x^{\prime}(z^{\prime})+{\rm i}y^{\prime}(z^{\prime})]=0$.
After the manipulation, the orbit can be described as
\begin{eqnarray}
r^{\prime}(z^{\prime})\exp\left[{\rm i}\theta^{\prime}(z^{\prime})\right]
=(1/2)\left[r\exp({\rm i}\theta)+{\rm i}\omega_{\beta}^{-1}
(p_{\perp}/\gamma_{\rm b}m_{\rm e})
\right. \nonumber \\ \left.
\times\exp({\rm i}\phi)\right]
\left[\exp(-{\rm i}\omega_{\beta}\zeta/v_{\rm b})
-\exp({\rm i}\omega_{\beta}\zeta/v_{\rm b})\right].
\label{eq:a7}
\end{eqnarray}
\noindent
As for the integrand of the RHS of equation~(\ref{eq:a5}),
we invoke the density inversion theorem (cf. Section~3.2),
which gives
\begin{equation}
\int_{0}^{\infty}{\rm d}U\frac{\partial}{\partial U}F_{\rm b}\left(H\right)=
\left[\frac{\partial\psi_{\rm b}(r)}{\partial r}\right]^{-1}
\frac{1}{{\hat n}_{\rm b}}\frac{\partial n_{\rm b}^{0}(r)}{\partial r}.
\label{eq:a8}
\end{equation}

Making use of equations~(\ref{eq:a7}) and (\ref{eq:a8}),
equation~(\ref{eq:a5}) (involving equation~\ref{eq:a6})
can be integrated, to finally yield
\begin{eqnarray}
\frac{\partial}{\partial r}\frac{1}{r}
\frac{\partial}{\partial r}r\delta A_{z}(r)
+\frac{4\pi{\rm i}\omega\sigma(r)}{c^{2}}\delta A_{z}(r)
\nonumber \\
=-\frac{4\pi e^{2}\beta_{\rm b}^{2}\Delta f_{\rm pair}}
{\gamma_{\rm b}m_{\rm e}}
\frac{\delta A_{z}(r)}{\left(\Omega^{2}-\omega_{\beta}^{2}\right)}
\frac{1}{r}\frac{\partial n_{\rm b}(r)}{\partial r}.
\label{eq:a9}
\end{eqnarray}
The form of equation~(\ref{eq:a9}) is similar to that of equation~(45)
in \citet{uhm80}, but in the RHS of the present equation~(\ref{eq:a9})
we have the corrections related to the pair production rate,
that is, the factor $\Delta f_{\rm pair}$ and the correction via
$\omega_{\beta}^{2}$ given in equations~(\ref{eq:12}) or (\ref{eq:22}).

\bsp

\label{lastpage}

\end{document}